\documentclass[twocolumn]{aastex63}

\received{30 Nov 2020}
\revised{19 Jan 2021}
\accepted{29 Jan 2021}

\submitjournal{ApJ Letters}

\shorttitle{Graphical Interpretation of Precipitation}
\shortauthors{Voit}

\begin{document}

\title{A Graphical Interpretation of Circumgalactic Precipitation}

\correspondingauthor{G. M. Voit}
\email{voit@msu.edu}

\author[0000-0002-3514-0383]{G. Mark Voit}
\affiliation{Michigan State University \\
Department of Physics and Astronomy \\
East Lansing, MI 48824, USA}



\begin{abstract}
Both observations and recent numerical simulations of the circumgalactic medium (CGM) support the hypothesis that a self-regulating feedback loop suspends the gas density of the ambient CGM close to the galaxy in a state with a ratio of cooling time to freefall time $\gtrsim 10$.  This limiting ratio is thought to arise because circumgalactic gas becomes increasingly susceptible to multiphase condensation as the ratio declines.  If the timescale ratio gets too small, then cold clouds precipitate out of the CGM, rain into the galaxy, and fuel energetic feedback that raises the ambient cooling time.  The astrophysical origin of this so-called precipitation limit is not simple but is critical to understanding the CGM and its role in galaxy evolution.  This paper therefore attempts to interpret its origin as simply as possible, relying mainly on conceptual reasoning and schematic diagrams.  It illustrates how the precipitation limit can depend on both the global configuration of a galactic atmosphere and the degree to which dynamical disturbances drive CGM perturbations.  It also frames some tests of the precipitation hypothesis that can be applied to both CGM observations and numerical simulations of galaxy evolution.
\end{abstract}


\keywords{miscellaneous --- galaxies}

\section{Introduction} 
\label{sec:intro}
 
This \textit{Letter} outlines some key concepts underlying the phenomenon sometimes called circumgalactic precipitation.  The topic's history stretches back to investigations of thermal instability in both stratified interstellar gas \citep[e.g.,][]{Spitzer_1956ApJ...124...20S,Field65,Defouw_1970ApJ...160..659D,bnf09} and the gaseous cores of galaxy clusters \cite[e.g.,][]{fn77,MathewsBregman_1978ApJ...224..308M,Cowie_1980MNRAS.191..399C,Nulsen_1986MNRAS.221..377N,Malagoli_1987ApJ...319..632M,bs89,Loewenstein_1989MNRAS.238...15L,McCourt+2012MNRAS.419.3319M,Gaspari+2013MNRAS.432.3401G,ChoudhurySharma_2016MNRAS.457.2554C,Voit_2017_BigPaper,Voit_2018ApJ...868..102V,Choudhury_2019MNRAS.488.3195C}.  Many of those rather technical papers are difficult for non-experts to interpret and yet have deep implications for both observations of the circumgalactic medium (CGM) and analyses of numerical simulations designed to model galaxy evolution.  

Several recent analyses of multiphase\footnote{The term \textit{multiphase} describes a medium in which temperatures and densities of neighboring regions differ by orders of magnitude.} circumgalactic gas produced in numerical simulations of galaxy evolution have underscored the need for a broader understanding of the astrophysics at play \citep[e.g.,][]{Lochhaas_2020MNRAS.493.1461L,Nelson_2020arXiv200509654N,Fielding_2020arXiv200616316F,Esmerian_2020arXiv200613945E}. Those simulations show that feedback from both supernovae and active galactic nuclei can maintain the CGM in a multiphase state characterized by radial profiles of pressure, gas density, temperature, and specific entropy that have a well-defined median.  Around each median profile, fluctuations in density, temperature, and entropy exhibit approximately log-normal distributions with long tails toward lower temperature and entropy and greater gas density.  A similar tail is generally not present in the pressure fluctuations, indicating that the tail consists of gas condensing out of the ambient medium and producing cooler clouds, also known as precipitation, in the CGM.

The explanatory illustrations presented here are therefore intended to help further an intuitive understanding of the CGM conditions that promote or inhibit development of a multiphase galactic atmosphere.  Section \ref{sec:damping} discusses thermal instability and how it damps in stratified galactic atmospheres.  Section \ref{sec:blob_orbits} shows how various perturbations of those atmospheres can overcome damping and produce multiphase condensation.  Section \ref{sec:tctff_pdf} relates the rate of multiphase condensation to both the median state of the galactic atmosphere and the dispersion of perturbations around the median. Section \ref{sec:tilting} briefly considers the consequences of altering a galactic atmosphere's entropy gradient. Section \ref{sec:summary} summarizes the paper.

\section{Damped Thermal Instability}
\label{sec:damping}

Thermal stability of a galactic atmosphere depends on whether a low-entropy perturbation within it cools more quickly (or heats more slowly) than its surroundings, causing its entropy contrast to increase with time \citep[e.g.,][]{Field65,Balbus86,Balbus88}.  In an optically thin medium, a low-entropy perturbation can grow because its greater particle density usually makes its radiative cooling rate greater.  If the background medium is homogeneous and not gravitationally stratified, then the entropy contrast of a low-entropy gas parcel continually increases, without damping, until it reaches a temperature that slows or halts radiative cooling.  Such a medium is thermally unstable and tends to develop multiphase structure on a cooling timescale
\begin{equation}
    t_{\rm cool} = \frac {3}{2} \frac {nkT} {n_e n_i \Lambda(T,Z,n)}
\end{equation}
where $\Lambda(T,Z,n)$ is the radiative cooling function of astrophysical plasma with temperature $T$, metallicity $Z$, and particle density $n$, defined with respect to electron density $n_e$ and ion density $n_i$.  

Gravity makes thermal instability more interesting.  The equilibrium configuration of a galactic atmosphere that is both hydrostatic and convectively stable in its potential well must have a negative pressure gradient and a positive entropy gradient.  Otherwise, convection sorts the gas parcels until specific entropy becomes a monotonically rising function of the gravitational potential $\phi$.  The result is an atmosphere with an entropy gradient that can be expressed in terms of the logarithmic slope $\alpha_K \equiv d \ln K / d \ln r$, where $K \equiv kTn_e^{-2/3}$, because changes in $\ln K$ are proportional to changes of specific entropy in a monatomic ideal gas.  

An adiabatic low-entropy perturbation in such an atmosphere does not remain a low-entropy perturbation because of buoyancy. Instead, it accelerates toward the center, causing its entropy contrast to decrease. Eventually it passes through a layer of equivalent specific entropy into a layer of lower specific entropy. It is then a high-entropy perturbation and begins to decelerate.  What follows is a series of buoyancy-driven oscillations with frequency
\begin{equation}
    \omega_{\rm buoy} \sim \alpha_K^{1/2} t_{\rm ff}^{-1}
    \; \; ,
\end{equation}
where the freefall time $t_{\rm ff} \equiv (2r/g)^{1/2}$ is based on the local gravitational acceleration $g$ \citep[e.g.,][]{Voit_2017_BigPaper}.  More formally, the perturbation has become an internal gravity wave with a Brunt-V\"ais\"al\"a frequency $\omega_{\rm buoy}$.

\begin{figure}[b]
\includegraphics[width=3.5in,trim=0.2in 0.0in 0.0in 0.0in]{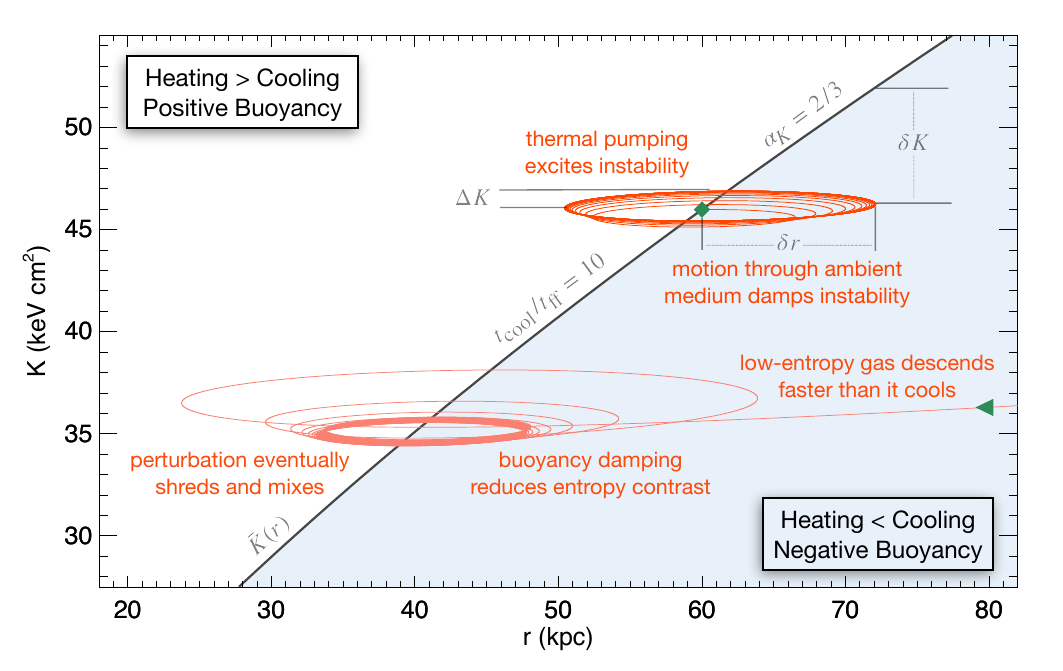}
\caption{Schematic illustrations of damped thermal instability in a stratified galactic atmosphere.  A thick charcoal line shows the entropy profile of a background atmosphere with $\alpha_K = 2/3$ and $t_{\rm cool} / t_{\rm ff} = 10$. In the blue region below that line, cooling exceeds heating, and gravity pulls the perturbation toward smaller $r$.  Above the line heating exceeds cooling, and buoyancy pushes the perturbation toward larger $r$.  The perturbation starting at the green diamond begins with a small amplitude and executes buoyant oscillations amplified by thermal pumping and damped by motion through the ambient medium.  Its amplitude ends up saturating with $\delta r / r \sim \alpha_K^{-1/2} (t_{\rm ff}/t_{\rm cool})$ and $\delta K / K \sim \alpha_K^{1/2} (t_{\rm ff} / t_{\rm cool})$.  The infalling perturbation starting on the lower right and marked by a green triangle begins with a larger amplitude and decays to the saturation amplitude as the perturbation sheds kinetic energy into the ambient medium.
\label{fig:damping}}
\end{figure}

Those buoyant oscillations can prevent thermal instability from producing multiphase structure in an otherwise static atmosphere with $\omega_{\rm buoy} t_{\rm cool} \gg 1$.  Figure~\ref{fig:damping} illustrates the reason.  The trajectories it shows were computed with the heuristic non-linear thermal instability model of \citet{Voit_2018ApJ...868..102V}.  At the upper right is a trajectory beginning at the green diamond on the atmosphere's median entropy profile, $\bar{K}(r)$. Along that median profile, heating equals cooling and $t_{\rm cool} / t_{\rm ff} = 10$. Initially, the perturbation is moving to greater altitude, into the part of the diagram where cooling exceeds heating.  Its entropy therefore starts to decline on a timescale $\sim t_{\rm cool}$, but it accelerates inward on a shorter timescale $\sim \omega_{\rm buoy}^{-1}$.  Before condensation can occur, the perturbation crosses the median entropy line, into the part of the diagram where heating exceeds cooling.  Its entropy then rises, it accelerates outward, and it returns to the median line.  Buoyant oscillations follow, as shown in the figure.  

While the perturbation remains small, thermal pumping amplifies those oscillations on a timescale $\sim t_{\rm cool}$.  Each time the perturbation goes below the median entropy, cooling drops it further below the median.  Each time it goes above the median, heating pushes it further above the median.  But amplitude growth also increases the damping rate of the oscillations, as motion through the ambient medium transfers increasing amounts of kinetic energy away from the perturbation.  Growth therefore saturates at an amplitude
\begin{equation}
    \frac {\delta r} {r} 
    \sim \left( \omega_{\rm buoy} t_{\rm cool} \right)^{-1}
    \sim \alpha_K^{-1/2} \frac {t_{\rm ff}} {t_{\rm cool}}
    \label{eq:saturation}
\end{equation}
at which energy losses to the ambient medium equal energy gains through thermal pumping \citep[e.g.,][]{Nulsen_1986MNRAS.221..377N,McCourt+2012MNRAS.419.3319M,Voit_2017_BigPaper}. 

\citet{Voit_2017_BigPaper} called this saturation process ``buoyancy damping" but misidentified the channel that drains kinetic energy from the perturbation.  The Erratum to \citet{Voit_2017_BigPaper} shows that internal gravity waves excited by thermal instability damp by coupling to resonant pairs of lower frequency gravity waves.  Those wave-triad interactions transfer kinetic energy away from the original unstable wave on a timescale $\sim [ \omega_{\rm buoy} (kr) (\delta r/r)]^{-1}$, where $k$ is the wavenumber of the original wave.  Growth of the original wave therefore saturates as this dissipation timescale approaches $t_{\rm cool}$. Long-wavelength disturbances ($kr \sim 1$) consequently saturate at the amplitude expressed in equation (\ref{eq:saturation}), corresponding to entropy fluctuations with a fractional amplitude $\delta K / K \sim \alpha_K ( \delta r / r ) \sim \alpha_K^{1/2} ( t_{\rm ff} / t_{\rm cool} )$, relative to the background.  However, the actual entropy changes experienced by a moving perturbation have an amplitude $\Delta K / K \sim (\omega_{\rm buoy} t_{\rm cool})^{-1} (\delta r/r) \sim \alpha_K^{-1} (t_{\rm ff} / t_{\rm cool} )^2$.

Perturbations starting with a larger amplitude than the saturation amplitude can also fail to result in multiphase condensation.  The lower trajectory in Figure~\ref{fig:damping} shows an example.  It enters the figure with $K \sim 0.6 \, \bar{K}$ at 80~kpc and proceeds toward smaller radii.  However, the perturbation it represents is falling faster than it can cool and ultimately descends through a layer of equivalent entropy at $r \approx 40$~kpc.  Buoyant oscillations of decaying amplitude follow until the perturbation reaches the saturation scale. A more complete model would include fluid instabilities that would shred the oscillating perturbation and mix it with the ambient medium, perhaps before it is able to reach the saturation scale. 

\section{Gravity Waves \& Precipitation}
\label{sec:blob_orbits}

Precipitation models have recently received considerable attention because both  observations \citep{Voit_2015Natur.519..203V,Voit+2015ApJ...803L..21V,Voit_2018ApJ...853...78V,Voit_2019ApJ...879L...1V,Voit_2019ApJ...880..139V,Hogan_2017_tctff,Pulido_2018ApJ...853..177P,Babyk_2018ApJ...862...39B} and simulations \citep{McCourt+2012MNRAS.419.3319M,Sharma_2012MNRAS.420.3174S,Gaspari+2012ApJ...746...94G,Gaspari+2013MNRAS.432.3401G,Li_2015ApJ...811...73L,Prasad_2015ApJ...811..108P,Prasad_2018ApJ...863...62P,YangReynolds_2016ApJ...818..181Y,Meece_2017ApJ...841..133M,Fielding_2017MNRAS.466.3810F,Esmerian_2020arXiv200613945E} suggest that coupling between energetic feedback and multiphase condensation enables at least some galactic atmospheres to self-regulate at a median ratio $t_{\rm cool} / t_{\rm ff} \approx 10$--20.  The precipitation hypothesis suggests that feedback fueled by cold, centrally accreting gas naturally suspends a galactic atmosphere in a state that is marginally unstable to multiphase condensation \citep[e.g.,][]{TaborBinney1993MNRAS.263..323T,BinneyTabor_1995MNRAS.276..663B,MallerBullock_2004MNRAS.355..694M,ps05,ps10,McCourt+2012MNRAS.419.3319M,Sharma_2012MNRAS.420.3174S,Gaspari+2012ApJ...746...94G,Gaspari_2017MNRAS.466..677G,Soker_2016NewAR..75....1S,Voit_2017_BigPaper}. However, the median $t_{\rm cool} / t_{\rm ff}$ ratio of the emergent marginal state is an order of magnitude greater than the value at which buoyancy should interfere with thermal instability and suppress multiphase condensation, according to the findings outlined in \S \ref{sec:damping}.  Something must therefore be offsetting the damping effects of buoyancy so that multiphase condensation can proceed.  This section outlines some of the possibilities.

\begin{figure}[t]
\includegraphics[width=3.5in,trim=0.2in 0.0in -0.1in 0.0in]
{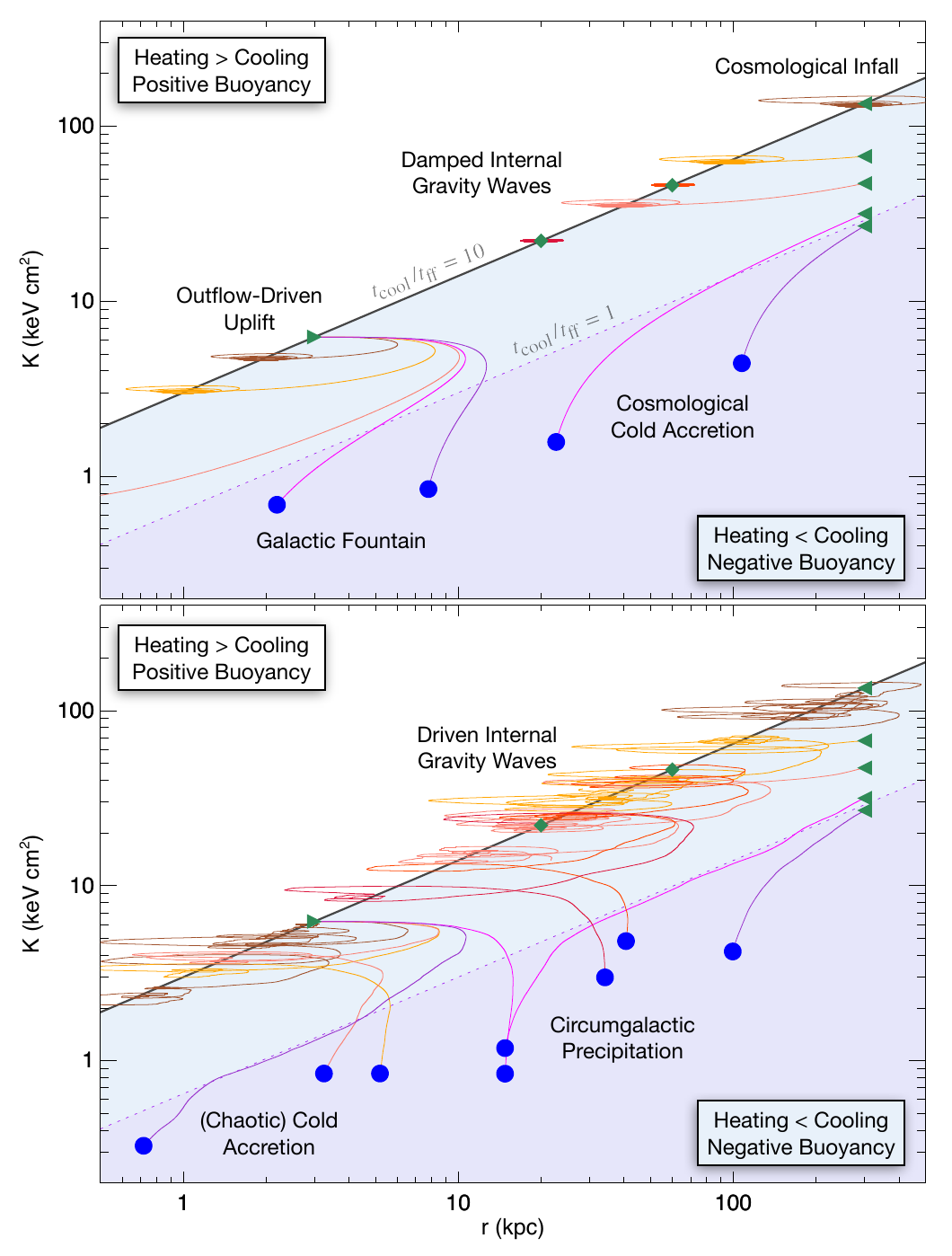}
\caption{Perturbation trajectories in idealized galactic atmospheres with a median entropy profile having $\alpha_K = 2/3$ and $t_{\rm cool} / t_{\rm ff} = 10$.  Lines and shading shared with Figure~\ref{fig:damping} have the same meanings.  Additionally, a dotted line traces where $t_{\rm cool} / t_{\rm ff} = 1$.  Perturbations entering the purple region below it inevitably condense, as symbolized by blue circles.  The upper panel shows trajectories computed with the model of \citet{Voit_2018ApJ...868..102V} for perturbations in a static atmosphere, and most of them converge to the saturation amplitude for thermally unstable but damped internal gravity waves.  Line colors for the Cosmological Infall and Outflow-Driven Uplift trajectories, in the order brown, orange, red, magenta, and purple, represent increasing susceptibility to condensation in a quiet atmosphere.  The lower panel shows what happens to those trajectories in a dynamically noisy atmosphere in which random momentum impulses buffet the perturbations, which can be considered internal gravity waves that have been driven to non-linear amplitudes.  Many of the trajectories then end in condensation, as described in \S \ref{sec:blob_orbits}.
\label{fig:blob_orbits}}
\end{figure}

The upper panel of Figure~\ref{fig:blob_orbits} illustrates two pathways that can lead to condensation in an otherwise static galactic atmosphere.  As in Figure~\ref{fig:damping}, all of the perturbation trajectories it shows were computed using the heuristic non-linear dynamical model from \citet{Voit_2018ApJ...868..102V}.  In fact, the trajectory beginning at 300~kpc and ending up at 40~kpc is identical to the infalling one in Figure~\ref{fig:damping}.  Several other infalling trajectories also begin at 300~kpc, and their fate depends on the perturbation's initial $t_{\rm cool} / t_{\rm ff}$ ratio.  If the ratio begins near unity, the perturbation can cool as least as quickly as it falls, allowing it to condense.  But if an infalling perturbation begins with a $t_{\rm cool} / t_{\rm ff}$ ratio much greater than unity, buoyancy damping causes it to settle into and merge with the ambient atmosphere at an entropy level not very different from its original value.  The fate of infalling gas coming from cosmological accretion therefore depends on both its initial  $t_{\rm cool} / t_{\rm ff}$ ratio and the entropy profile of the galactic atmosphere it is entering. A complementary analysis of this perturbation mode can be found in \citet{Choudhury_2019MNRAS.488.3195C}, who consider how development of a multiphase medium depends on the amplitude of isobaric density perturbations and find that condensation of those perturbations depends jointly on their initial amplitude and the ambient value of $t_{\rm cool}/t_{\rm ff}$.

The second condensation pathway in the upper panel of Figure~\ref{fig:blob_orbits} begins at small radii.  Outflows that lift low-entropy gas to greater altitude can stimulate condensation if they are able to make $t_{\rm cool} / t_{\rm ff} \lesssim 1$ in the uplifted gas \citep[e.g.,][]{Revaz_2008A&A...477L..33R,LiBryan2014ApJ...789..153L,McNamara_2016ApJ...830...79M,Voit_2017_BigPaper}.  In the atmosphere pictured, uplift by a factor $\sim 3$ in radius is required.  Atmospheres with greater median values of $t_{\rm cool} / t_{\rm ff}$ require greater amounts of uplift.  The essential features of this condensation mode were captured decades ago by galactic fountain models for the origin of high-velocity clouds in a galactic atmosphere \citep{ShapiroField_1976ApJ...205..762S,Bregman_GalacticFountain_1980ApJ...236..577B} and are inherent in modern simulations of condensing galactic winds \citep[e.g.,][]{Vijayan_2018,Schneider_2018ApJ...862...56S}. 

Trajectories in the lower panel of Figure~\ref{fig:blob_orbits} begin with the same initial conditions as those in the upper panel.  The only difference is the presence of dynamical noise in the ambient atmosphere.  Perturbation trajectories in the \citet{Voit_2018ApJ...868..102V} model can be given random momentum impulses intended to resemble the effects of turbulence and other forms of kinetic disturbance.  That feature of the model was inspired by the finding of \citet{Gaspari+2013MNRAS.432.3401G} that driving of turbulence in a galactic atmosphere with a median ratio $t_{\rm cool} / t_{\rm ff} \approx 10$ can interfere with buoyancy and promote multiphase condensation. The perturbations that condense correspond to internal gravity waves that have been driven toward amplitudes large enough for the perturbation's \textit{local} value of $t_{\rm cool} / t_{\rm ff}$ to approach unity.

The lower panel of Figure~\ref{fig:blob_orbits} shows that dynamical noise can cause condensation of perturbations that would otherwise damp and converge to the saturation amplitude.  Two examples are the red trajectories starting at the green diamonds located 20~kpc and 60~kpc from the center.  Those trajectories represent internal gravity waves that saturate in the upper panel but result in condensation near $r \approx 40$~kpc in the bottom panel.  Two more examples are the orange and salmon colored trajectories starting 3~kpc from the center.  In the upper panel, uplift alone is not enough to make them condense, but random momentum impulses in the lower panel end up driving those trajectories into condensation near $r \approx 4$~kpc. 

Collectively, all of the condensing trajectories in the lower panel of Figure~\ref{fig:blob_orbits} can be considered routes to precipitation.  Grouping all of them into a generic category called ``precipitation" is useful, because it can be difficult to infer the origin of condensing CGM gas from the location where it condenses.  Consider, for example, the two trajectories that end in condensation near $r \approx 15$~kpc, which are represented with magenta lines.  One comes from infall of a gas parcel with $t_{\rm cool} / t_{\rm ff} \approx 1$.  The other comes from uplifted gas originally at $r \approx 3$~kpc.  In a static atmosphere, the uplifted gas parcel condenses at $r \approx 2$~kpc (see upper panel).  But in the noisy atmosphere of the lower panel, random momentum impulses push that parcel's trajectory considerably farther from the center, leading to condensation in the CGM.  

Notice also that random momentum impulses can push other condensing gas parcels inward. In the lower panel, the uplifted parcel that initially has the greatest outward velocity at 3~kpc (purple line) ends up condensing closer to the center, at $r < 1$~kpc.  This form of precipitation near the center of a galaxy is representative of the chaotic cold accretion process that can strongly boost the feedback output from a massive galaxy's central black hole \citep[e.g.,][]{ps05,Gaspari+2013MNRAS.432.3401G,Tremblay_2016Natur.534..218T}. 

\section{The Global Precipitation Limit}
\label{sec:tctff_pdf}

\begin{figure}[t]
\includegraphics[width=3.5in,trim=0.2in 0.0in -0.1in 0.0in]{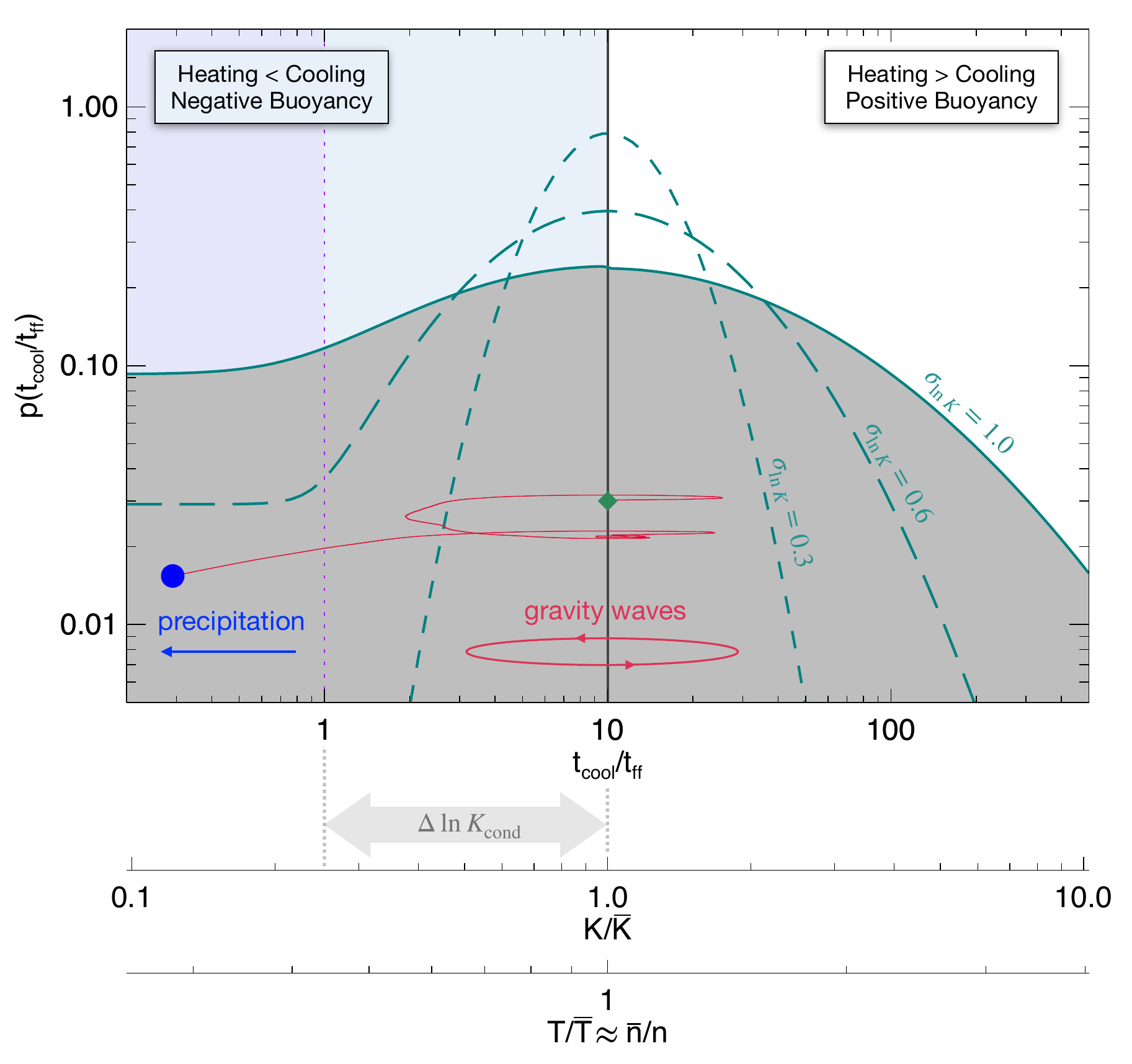}
\caption{Schematic probability distributions of $t_{\rm cool} / t_{\rm ff}$, $K/\bar{K}$, $T/\bar{T}$, and $\bar{n}/n$ for precipitating atmospheres in approximate pressure balance.  The cores of the distributions are log-normal, with $\sigma_{\ln K} = 1.0$ (solid line), 0.6 (long-dashed line), and 0.3 (short-dashed line), and the two with larger dispersion have a flat tail below $t_{\rm cool} / t_{\rm ff} = 1$.  In the core of each distribution, buoyancy causes entropy perturbations to oscillate around the median as gravity waves.  But in the tail, cooling operates faster than buoyancy, allowing perturbations that reach the tail to proceed into multiphase condensation.  The thin line beginning at the green diamond and ending at the blue circle shows an example.  Given such a distribution, the atmosphere's precipitation rate is proportional to the height of the tail and depends on how $\sigma_{\ln K}$ compares with the fractional difference $ \Delta \ln K_{\rm cond}$ between $\bar{K}$ and the local entropy level at which precipitation can occur. 
\label{fig:tctff_pdf}}
\end{figure}

The trajectories in Figure~\ref{fig:blob_orbits} demonstrate that $t_{\rm cool} / t_{\rm ff} \lesssim 1$ is a local criterion for precipitation in a stratified atmosphere.\footnote{The criterion $\omega_{\rm buoy} t_{\rm cool} \lesssim 1$ is more general (see \S \ref{sec:tilting})}  Individually, those trajectories do not account for why a galactic atmosphere would self-regulate near a much larger median ratio $t_{\rm cool} / t_{\rm ff} \sim 10$, but collectively they provide a clue.  Internal gravity waves driven by dynamical noise in the lower panel's calculations typically reach amplitudes roughly half of what is necessary to end in condensation.  The distribution of those amplitudes can be characterized by the resulting dispersion $\sigma_{\ln K}$ of the fractional entropy perturbation amplitude $\delta \ln K = [K - \bar{K}(r)] / \bar{K}(r)$ relative to the local median.  The low-entropy tail of that distribution represents perturbations transitioning into condensation.  Consequently, the precipitation rate of an atmosphere depends on how its typical perturbation amplitude $\sigma_{\ln K}$ compares with the fractional difference $ \Delta \ln K_{\rm cond} = | K - \bar{K} | / \bar{K}$ between the median entropy $\bar{K}$ and the entropy level $K$ at which $t_{\rm cool} / t_{\rm ff} \approx 1$ within a perturbation (see Figure~\ref{fig:tctff_pdf}). That link between median entropy and precipitation rate enables a galactic atmosphere to self-adjust so that feedback fueled by precipitation deposits just enough thermal energy in the CGM to balance radiative cooling, as outlined in \citet{Voit_2018ApJ...868..102V}.

Figure~\ref{fig:tctff_pdf} presents the idea schematically.  It shows distribution functions that are log-normal around a median entropy $\bar{K}$, except for a flat tail at the low-entropy end, where $t_{\rm cool}/t_{\rm ff} \lesssim 1$.  The figure presents a case in which the median entropy $\bar{K}$ corresponds to $t_{\rm cool} / t_{\rm ff} = 10$. More generally, $\bar{K}$ can be considered an adjustable parameter of the distribution function that determines the median $t_{\rm cool}/t_{\rm ff}$ ratio.  The relationship shown between $t_{\rm cool}/t_{\rm ff}$ and the ratio $K/\bar{K}$ assumes pressure balance and $\Lambda \propto T^{-0.8}$, which is appropriate for $10^{5.5} \, {\rm K} < T < 10^{6.5} \, {\rm K}$ \citep[e.g.,][]{sd93,Schure_2009A&A...508..751S}.  Together, those assumptions give $t_{\rm cool} \propto K^{1.7}$ and the relationships $\sigma_{\ln t_{\rm cool}} \approx 1.7 \, \sigma_{\ln K}$ and $\sigma_{\ln t_{\rm cool}} \approx 2.8  \, \sigma_{\ln T}$ \citep[e.g.,][]{Voit_2019ApJ...880..139V}.

Evidence for such a perturbation distribution can be observed in both idealized and cosmological simulations of feedback-regulated circumgalactic gas.  See, for example, Figure~1 of \citet{Lochhaas_2020MNRAS.493.1461L} and especially Figure~4 of \citet{Fielding_2020arXiv200616316F}.  The distributions of $T/\bar{T}$ appear log-normal in the core and have tails that level off near $0.3 \, T/\bar{T}$.  \citet{Fielding_2020arXiv200616316F} show that the dispersions of those distributions differ, depending on features peculiar to each simulation.  The cosmological simulations \citep[from][]{Joung_Accretion_2012ApJ...759..137J, Springel_TNG_2018MNRAS.475..676S} analyzed by \citet{Fielding_2020arXiv200616316F} show greater dispersion in CGM conditions, particularly at large radii, presumably because they include dynamical disturbances, such as cosmological infall, mergers, and stirring of the CGM by orbiting subhalos, that are not present in idealized simulations.  In those cosmological simulations, the peak of the log-normal distribution is roughly twice the level of the flat tail, similar to the distribution with $\sigma_{\ln K} = 1$ in Figure~\ref{fig:tctff_pdf}.  Interestingly, the corresponding temperature dispersion ($\sigma_{\ln T} \approx 0.6$) is consistent with the one inferred by \citet{Voit_2019ApJ...880..139V} from observations of CGM O~VI absorption around galaxies like the Milky Way.

For the idealized distribution functions in Figure~\ref{fig:tctff_pdf}, the precipitation rate is exponentially sensitive to the ratio $\Delta \ln K_{\rm cond} / \sigma_{\ln K}$, because
\begin{equation}
    p \left( \frac {t_{\rm cool}} {t_{\rm ff}} \right) 
     \: \propto \: 
      \exp \left[ - \frac {1} {2} 
             \left( \frac { \Delta \ln K_{\rm cond}} 
                          {\sigma_{\ln K}} \right)^2 \right]
    \label{eq:pdf}
\end{equation}
for $t_{\rm cool} / t_{\rm ff} \gtrsim 1$. The model represented in the figure has $\Delta \ln K_{\rm cond} = (\ln 10) / 1.7 \approx 1.35$, because $t_{\rm cool} / t_{\rm ff} = 10$ at $\bar{K}$ and $t_{\rm cool} \propto K^{1.7}$.  Given this difference between $\bar{K}$ and the value of $K$ at which $t_{\rm cool} / t_{\rm ff} = 1$, approximately 9\% of a log-normal distribution with $\sigma_{\ln K} = 1.0$ has $t_{\rm cool} / t_{\rm ff} \lesssim 1$. That percentage drops to $\sim 1$\% in the distribution with $\sigma_{\ln K} = 0.6$.  Precipitation depletes the fraction of the ambient atmosphere with $t_{\rm cool} / t_{\rm ff} \lesssim 1$ on a dynamical timescale (because $t_{\rm cool} \approx t_{\rm ff}$ in those perturbations).  And dynamical noise presumably restores the tail of the distribution function on a similar timescale.  

That line of reasoning leads to a hypothesis about the precipitation rate: In a given layer of a galactic atmosphere it should be similar to the gas mass of the layer, divided by $t_{\rm ff}$, times the fraction of mass with $t_{\rm cool} / t_{\rm ff} \lesssim 1$.  Within the virial radius, the expected precipitation rate of a galactic atmosphere represented by the solid line in Figure~\ref{fig:tctff_pdf} ($\sigma_{\ln K} = 1.0$) is then similar to the CGM gas mass divided by a Hubble time.  The precipitation rate corresponding to the long-dashed line ($\sigma_{\ln K} = 0.6$) is an order of magnitude smaller.  In a state of long-term balance, entropy fluctuations cannot be much greater than $\sigma_{\ln K} = 1.0$ without exhausting the ambient CGM (unless the median ratio is $t_{\rm cool} / t_{\rm ff} \gg 10$), and they cannot be much smaller than $\sigma_{\ln K} = 0.6$ while still providing enough precipitation to fuel feedback. This hypothesis for the CGM precipitation rate seems reasonable but remains to be tested with numerical simulations.

Another observable CGM property connected with the distribution functions in Figure~\ref{fig:tctff_pdf} is the velocity dispersion of internal gravity waves.   According to the numerical experiments in \citet{Voit_2018ApJ...868..102V}, 
the dynamical disturbances required to promote precipitation in an atmosphere with $\alpha_K \approx 2/3$ and a median ratio $t_{\rm cool} / t_{\rm ff} \sim 10$--20 maintain a one-dimensional ``turbulent" velocity dispersion $\sigma_{\rm t} \approx 0.3$--0.4$ \, v_{\rm c}$.  Note that the density fluctuations shown in Figure \ref{fig:tctff_pdf} can be considerably larger than the compressive density fluctuations produced by turbulent speeds of this magnitude in a hydrostatic atmosphere, which have fractional amplitudes $\delta \ln n \sim (\sigma_{\rm t}/v_{\rm c})^2$.  That is because the fluctuations depicted in Figure \ref{fig:tctff_pdf} arise instead from vertical displacements of gas in an atmosphere with an entropy gradient. Recent numerical simulations by \citet{Mohapatra_2020,Mohapatra_2021} have shown turbulence in a stratified medium generates density fluctuations of amplitude $\delta \ln n \sim \alpha_K (\sigma_{\rm t}/v_{\rm c})$.  

The critical velocity dispersion implied by the \citet{Voit_2018ApJ...868..102V} model is broadly consistent with the findings of \citet{Gaspari+2013MNRAS.432.3401G} for numerical simulations that drive turbulence, as well as with the observed velocity dispersions of galaxy-cluster cores that appear to be precipitating \citep{Gaspari_2018ApJ...854..167G}.  The turbulence cannot be much greater without either damping the perturbations through mixing with the ambient gas or overheating the ambient medium through turbulent dissipation \citep{Gaspari+2013MNRAS.432.3401G,Gaspari_2017MNRAS.466..677G,BanerjeeSharma_2014MNRAS.443..687B,Buie_2018ApJ...864..114B,Buie_stratified_2020ApJ...896..136B}.  Around a galaxy like the Milky Way, the predicted one-dimensional velocity dispersion of hot gas in a precipitating CGM is therefore $\sigma_{\rm t} \approx 50$--$70 \, {\rm km \,s^{-1}}$.  Around a massive elliptical galaxy the prediction rises to $\sigma_{\rm t} \approx 100$--$150 \, {\rm km \,s^{-1}}$.

\section{Tilting of the Entropy Profile}
\label{sec:tilting}

So far, the paper has focused entirely on entropy profiles with a constant median $t_{\rm cool} / t_{\rm ff}$ ratio, corresponding to $\alpha_K = 2/3$ in an isothermal potential well.  Tilting the entropy profile away from that special slope has consequences.  The most extreme case is an isentropic atmosphere with $\alpha_K = 0$, which eliminates buoyancy and produces singularities in the model of \S \ref{sec:damping}.  In that case, buoyancy is unable to damp thermal instability, allowing low-entropy perturbations to condense on a timescale $\sim t_{\rm cool}$ regardless of the median $t_{\rm cool} / t_{\rm ff}$ ratio.  This may be how massive galaxies in the Illustris TNG50 simulation manage to maintain multiphase circumgalactic gas, even though their ambient atmospheres have a median $t_{\rm cool} / t_{\rm ff} \sim 30$--100 at $r \sim 30$--300~kpc \citep{Nelson_2020arXiv200509654N}. Those large values of $t_{\rm \rm cool} / t_{\rm ff}$ would appear to be unfavorable to development of a multiphase medium, but the median entropy slope in that radial range is only $\alpha_K \approx 1/4$ in this subsample of massive galaxies from TNG50. Also, many of the individual entropy profiles have radial intervals with $\alpha_K \lesssim 0$, making those regions convectively unstable, free from buoyancy damping, and susceptible to precipitation.

Tilting the entropy slope the opposite way has consequences that may be more dramatic.  An entropy profile steep enough for the median $t_{\rm cool} / t_{\rm ff}$ ratio to rise with radius tends to focus multiphase condensation onto the galaxy's center, potentially supercharging feedback from the galaxy's central black hole. \citet{Voit_valve_2020arXiv200609381V} have recently demonstrated how such a tilt may link quenching of star formation by black-hole feedback with a galaxy's central stellar velocity dispersion.  Observations indeed show that central entropy profiles with $\alpha_K \gg 1$ are rare \citep[e.g.,][]{Lakhchaura_2018MNRAS.481.4472L,Babyk_2018ApJ...862...39B}.  And observations of some massive elliptical galaxies with $\alpha_K \approx 1$ show that this slope flattens within the central kiloparsec \citep{Werner+2012MNRAS.425.2731W,Frisbie_2020arXiv200612568F}, allowing centralized precipitation that is maximized at radii where $\omega_{\rm buoy} t_{\rm cool}$ is minimized \citep{Voit+2015ApJ...803L..21V}.

\section{Summary}
\label{sec:summary}

This paper has attempted to present the following key concepts of self-regulating circumgalactic precipitation as simply as possible, primarily through schematic diagrams.
\begin{enumerate}

    \item \textit{Buoyancy Damping.}  Thermal instability in static galactic atmospheres with $\alpha_K^{1/2} (t_{\rm cool} / t_{\rm ff}) \approx \omega_{\rm buoy} t_{\rm cool} \gg 1$ drives internal gravity-wave oscillations that damp before producing multiphase condensation (see \S \ref{sec:damping} and Figure \ref{fig:damping}). Instead of condensing, those thermally unstable perturbations saturate at a fractional entropy amplitude $\delta \ln K \sim \alpha_K^{1/2} (t_{\rm ff}/t_{\rm cool})$, relative to the background medium.
    
    \item \textit{Local Precipitation Threshold.}  Because of buoyancy damping, a low-entropy perturbation within an atmosphere having $\alpha_K^{1/2} (t_{\rm cool} / t_{\rm ff}) \approx \omega_{\rm buoy} t_{\rm cool} \gg 1$ must have a local cooling time satisfying the criterion $\omega_{\rm buoy} t_{\rm cool} \lesssim 1$ in order to condense (see \S \ref{sec:tctff_pdf} and Figure \ref{fig:blob_orbits}).  If the atmosphere's entropy slope is $\alpha_K \approx 2/3$, then the local threshold for precipitation is $t_{\rm cool} / t_{\rm ff} \lesssim 1$.
    
    \item \textit{Uplift.}  Galactic outflows that lift ambient gas nearly adiabatically can lower the local $t_{\rm cool} / t_{\rm ff}$ ratio of the uplifted gas by increasing $t_{\rm ff}$ without significantly changing $t_{\rm cool}$ (see Figure \ref{fig:blob_orbits}, upper panel).  The amount of uplift required to achieve $t_{\rm cool} / t_{\rm ff} \lesssim 1$ depends on the global median $t_{\rm cool} / t_{\rm ff}$ ratio.  Therefore, the global ratio is a measure of the atmosphere's susceptibility to multiphase condensation, when it is disturbed.  This route to condensation resembles a classic galactic fountain.  
    
    \item \textit{Infall.} Alternatively, cosmological infall or stripping of low-entropy gas from an orbiting subhalo can introduce perturbations of non-linear amplitude that are able to condense if they start with $\omega_{\rm buoy} t_{\rm cool} \lesssim 1$ (see Figure \ref{fig:blob_orbits}, upper panel).  \item \textit{Dynamical Driving.}  Hydrodynamical disturbances strong enough to interfere with buoyancy damping produce additional opportunities for condensation (see Figure \ref{fig:blob_orbits}, lower panel).  If all CGM perturbations are represented as internal gravity waves of amplitude $\delta \ln K \approx \alpha_K (\delta \ln r)$, then condensation corresponds to driving of those gravity-wave oscillations to amplitudes that locally satisfy $\omega_{\rm buoy} t_{\rm cool} \lesssim 1$.  Drivers of CGM fluctuations may include galactic winds, turbulence, cosmological infall, or stirring by orbiting subhalos.
    
    \item \textit{Circumgalactic Precipitation.}  Given the randomness produced by multiple sources of dynamical driving, it may be difficult to pinpoint the origin of a particular multiphase gas cloud in the CGM (see Figure \ref{fig:blob_orbits}, lower panel).  However, considering all routes to condensation to be forms of circumgalatic precipitation helps to link their collective presence with the global characteristics of the ambient galactic atmosphere.
    
    \item \textit{Global Precipitation Limit.}  The susceptibility to precipitation of a stratified galactic atmosphere ($\alpha_K \sim 1$) depends on both its median $t_{\rm cool} / t_{\rm ff}$ ratio and the dispersion $\sigma_{\ln K}$ of entropy fluctuations within it (see \S \ref{sec:tctff_pdf} and Figure \ref{fig:tctff_pdf}).  The median $t_{\rm cool} / t_{\rm ff}$ ratio determines the fractional entropy difference $\Delta \ln K_{\rm cond}$ between the atmosphere's median entropy $\bar{K}$ and the entropy of a perturbation in which $t_{\rm cool} / t_{\rm ff} \lesssim 1$.  The fraction of the CGM that is able to precipitate therefore depends on the ratio $\Delta \ln K_{\rm cond} / \sigma_{\ln K}$.  Numerical simulations of the CGM indicate that its entropy fluctuations have a log-normal distribution around the median, with $\sigma_{\ln K} \approx 0.6$--1.0, implying that the CGM precipitation rate is exponentially sensitive to $\Delta \ln K_{\rm cond} / \sigma_{\ln K}$.  Consequently, the ratio $f \equiv \Delta \ln K_{\rm cond} / \sigma_{\ln K}$ needs to be large enough to avert catastrophic precipitation and an overwhelming feedback response.  In a typical galactic atmosphere (with $t_{\rm cool} \propto K^{1.7}$), these considerations yield a global precipitation limit
    \begin{equation}
        t_{\rm cool} / t_{\rm ff} 
          \: \gtrsim \: \exp \left( 1.7 f \sigma_{\rm ln K} \right)
        \label{eq:global_limit}
    \end{equation}  
    on the median ratio that reduces to $t_{\rm cool} / t_{\rm ff} \gtrsim 10$ for $f \sigma_{\ln K} \approx 1.35$.  Self-regulating precipitation converges to a value of $f$ at which accretion of cold gas into the galaxy fuels just enough  feedback to keep the CGM in approximate thermal balance.  That equilibrium value is likely to be in the range $1 \lesssim f \lesssim 2$ because of the exponential sensitivity of the precipitation criterion. 
    
    \item \textit{Observable Features.}  Two distinct observable features allow tests of this interpretation of circumgalactic precipitation.  Entropy fluctuations in approximate pressure balance with their surroundings correspond to temperature and density fluctuations with log-normal dispersion $\sigma_{\ln T} \approx \sigma_{\ln n} \approx 0.6 \, \sigma_{\rm ln K}$ (see Figure \ref{fig:tctff_pdf}).  Around a galaxy like the Milky Way, with a CGM temperature $\sim 10^6 \, {\rm K}$, the O~VI absorption-line column densities are sensitive to the amplitude of temperature fluctuations, and observations are so far consistent with $\sigma_{\ln T} \sim 0.6$ \citep{Voit_2019ApJ...880..139V}.  Also, dynamical disturbances great enough to cause precipitation maintain a radial velocity dispersion $\approx 0.3$--$0.4 \, v_{\rm c}$ in the CGM gas, amounting to 50--$70 \, {\rm km \, s^{-1}}$ around a galaxy like the Milky Way and 100--$150 \, {\rm km \, s^{-1}}$ around a massive elliptical galaxy.  However, additional modeling will be required to obtain predictions of correlations between O~VI column density and absorption-line width.
    
    \item \textit{Entropy-Profile Tilt.}  All of these predictions for precipitation substantially change if the CGM entropy profile becomes nearly flat ($\alpha_K \ll 1$) because buoyancy damping is eliminated as $\alpha_K \rightarrow 0$. All low-entropy perturbations can then condense on a timescale $\sim t_{\rm cool}$, regardless of the median $t_{\rm cool} / t_{\rm ff}$ ratio (\S \ref{sec:tilting}).  Those atmospheres call for an amendment to equation (\ref{eq:global_limit}) that replaces $t_{\rm cool} / t_{\rm ff}$ with the atmosphere's median value of $\omega_{\rm buoy} t_{\rm cool}$, giving a more general global precipitation limit $\omega_{\rm buoy} t_{\rm cool} \gtrsim 10$ for $f \sigma_{\ln K} \approx 1.35$. 
     
     \newpage
     
    \item \textit{Minimum Precipitation Limit.}  Notice also that thermal instability near the saturation limit can generate perturbations large enough to precipitate without dynamical driving, as along as the median of $t_{\rm cool}/t_{\rm ff}$ is sufficiently small.  Making another change to equation (\ref{eq:global_limit}) by replacing $\sigma_{\ln K}$ with the saturation amplitude $\delta \ln K \sim \alpha_K (\omega_{\rm buoy} t_{\rm cool})^{-1}$ gives the relation
    \begin{equation}
      \omega_{\rm buoy} t_{\rm cool} \cdot  
        \ln ( \omega_{\rm buoy} t_{\rm cool} ) 
          \: \gtrsim \: 1.7 f \alpha_K    
    \end{equation}
    which implies a firm lower limit on the median ratio of $t_{\rm cool} / t_{\rm ff} \gtrsim 2.5$ for $f \alpha_K \sim 1$. 
\end{enumerate}

\acknowledgements{
This work was supported in part by grant TM8-19006X from the Chandra Science Center.  Comments from Ed Buie, Megan Donahue, Rachel Frisbie, Forrest Glines, Claire Kopenhafer, and Deovrat Prasad helped to improve its clarity.
}





\bibliographystyle{aasjournal}

\end{document}